\documentclass[twocolumn,showpacs,preprintnumbers,amsmath,amssymb]{revtex4}

\usepackage{graphicx}
\usepackage{dcolumn}
\usepackage{bm}

\begin{document}


\title{Anomalous Flux Flow Resistivity in Two Gap Superconductor  MgB$_2$ }

\author{A.~Shibata, M.~Matsumoto, K.~Izawa, and  Y.~Matsuda}
\affiliation{Institute for Solid State Physics, University of Tokyo, Kashiwanoha 5-1-5, Kashiwa, Chiba 277-8581, Japan}%
\author{S.~Lee,  and S.~Tajima}
\affiliation{ Superconductivity Research Laboratory, ISTEC, 1-10-13 Shinonome, Koto-ku, Tokyo 135-0062, Japan }%

\date{\today}

\begin{abstract}

The flux flow resistivity $\rho_f$  associated with  purely viscous motion of  vortices in high-quality MgB$_2$ was measured by  microwave surface impedance.   Flux flow resistivity exhibits unusual field dependence with strong enhancement at low field, which is markedly different to conventional $s$-wave superconductors.    A crossover field which separates two distinct flux flow regimes having different $\rho_f$ slopes was clearly observed in {\boldmath $H$}$\parallel ab$-plane.    The unusual $H$-dependence indicates that  two very differently sized superconducting gaps in MgB$_2$ manifest  in the vortex dynamics and  almost equally contribute to  energy dissipation.  The carrier scattering rate in two different bands is also discussed with the present results, compared to heat capacity and thermal conductivity results.

\end{abstract}


\maketitle

Since the recent discovery of  high temperature superconductivity in MgB$_2$, great attention has been directed towards understanding the detailed nature of superconductivity \cite{akimitsu,can}.      MgB$_2$ consists of two bands having roughly equal density of states (DOS),  strongly two dimensional $\sigma$-band ($sp_xp_y$) and three dimensional $\pi$-band  ($p_z$) bands.   Associated with these two bands, two very differently sized superconducting gaps  are quite distinct and manifest themselves clearly in many physical properties, which have become one of the central topic of the physical properties of MgB$_2$.   In fact, apart from the much debated pairing mechanism of high transition temperature superconductivity,  physical properties in the superconducting state  {\it i.e.}, the distinguishing characteristics of a superconductor with two gaps, are still largely unexplored.     Recent  scanning tunneling microscope (STM) \cite{stm}, heat capacity \cite{hc}, and thermal conductivity \cite{tc}
 measurements have revealed that the physical properties in the vortex state of MgB$_2$ are dramatically different from those expected in conventional $s$-wave superconductors.     Despite these extensive studies, however, the dynamical properties of the vortices, such as  the energy dissipation associated with the vortex motion which is intimately related to the electronic structure \cite{kopnin,esc,kambe,matsuda},  are still far from being completely understood.     
	
	   When a vortex line in a type-II superconductor responds to a driving current,   frictional force is given by the damping viscosity, which in turn depends on the energy dissipation process of quasiparticles in and around the vortex cores.   To gain an understanding of this energy dissipation, experimental determination of the free flux flow (FFF) resistivity is particularly important.    FFF  refers to a purely viscous vortex motion of vortices, which is realized when the vortices move in pinned free states.    The FFF state in conventional $s$-wave superconductors has been extensively studied and a rather good understanding has been developed.   In $s$-wave superconductors,  quasiparticles trapped within the vortex core that form  Caroli-de~Gennes-Matricon bound states play a key role in the dissipation process \cite{kopnin,esc}.    The FFF resistivity $\rho_f$ is essentially proportional to normal state resistivity $\rho_n$ and applied field,
\begin{equation}
\rho_f=\frac{H}{H_{c2} }\rho_n.
\label{eqn:FFFS}
\end{equation}
This Bardeen-Stephen relationship has been confirmed in most  diry  and clean $s$-wave systems throughout almost the whole Abrikosov phase.  Recently,  striking deviation of  FFF resistivity from Eq.(1) has been reported in unconventional superconductors with gap nodes \cite{kambe,matsuda,takagi,izawa,tsuchiya,mat}.   For instance, in $d$-wave Bi$_2$Sr$_2$CuO$_{6+\delta}$,  $\rho_f$ exhibits a $H$-linear dependence at low fields, followed by a non-linear $H$-dependence described well by,
\begin{equation}
\rho_f \propto \sqrt{\frac{H}{H_{c2} }}\rho_n,  
\label{eqn:FFFD}
\end{equation}
at higher fields \cite{matsuda}.  This demonstrates that the energy dissipation is strongly influenced by the superconducting gap structure.  However, despite these extensive studies of the FFF state, the detailed microscopic mechanism of energy dissipation is not well understood, exposing our  incomplete knowledge of vortex dynamics.  This situation thus calls for a new textbook example of  FFF resistivity in a superconductor with a distinctive gap structure.  The FFF resistivity of MgB$_2$ with two distinct superconducting gaps may provide an unique opportunity for studying how the energy dissipation occurs when vortices move in the superfluid. 

	In this paper,   FFF resistivities of  MgB$_2$ are examined by measuring microwave surface impedance.   FFF resistivity  exhibits unusual field dependence, which is markedly different from conventional $s$-wave superconductors.   These results are discussed in terms of  energy dissipation in the superconductors with two distinct gaps.
		
	High quality single crystals of MgB$_2$ ($T_c$=38.6~K) were grown using a high pressure method \cite{lee}.  Typical  size was 0.3 $\times$ 0.3 $\times$ 0.1~mm$^3$.   Microwave surface impedance $Z_s=R_s+iX_s$, where $R_s$ and $X_s$ are the surface resistance and surface reactance, respectively, was measured by the standard cavity perturbation technique \cite{klein}.   Cylindrical cavity resonators made of Cu ($Q\simeq$44000) and superconducting Pb ($Q\simeq 1.0\times10^6$)  operated at 28.5~GHz were used  in  TE$_{011}$ mode.   Sample was placed in an antinode of the oscillatory magnetic field {\boldmath $H$}$_{\omega}$.    $Z_s$ was measured in two different configurations,  {\boldmath $H$}$_{\omega}\parallel ${\boldmath $H$}$\parallel c$-axis $(Z_s^{H_{\omega}\parallel c})$ and  {\boldmath $H$}$_{\omega}\parallel ${\boldmath $H$}$\parallel ab$-plane  $(Z_s^{H_{\omega}\parallel ab})$.   
	
\begin{figure}[b]
\begin{center}
\includegraphics[width=3in]{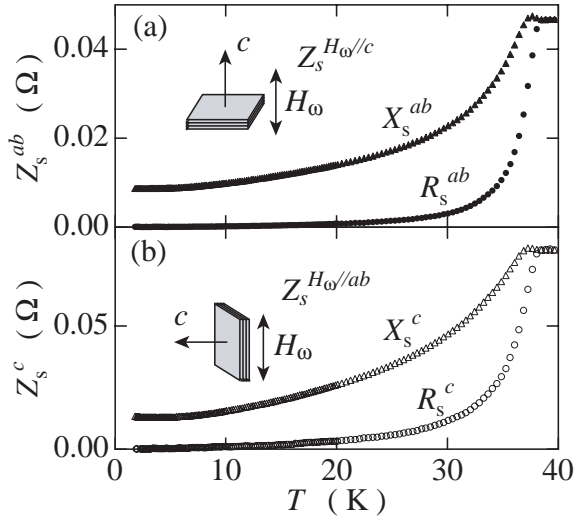}
\caption{ (a) Temperature dependence of  in-plane  surface impedance, $Z_s^{ab}=R_s^{ab}+iX_s^{ab}$, in MgB$_2$ single crystal.    (b)The same data for out-of-plane surface impedance, $Z_s^c=R_s^c+iX_s^c$.   $\rho_n^{ab}=2\mu \Omega$ cm, $\rho_n^{c}/\rho_n^{ab}=3$ and $S^{ab}=3S^c$ were used.}
\end{center}
\end{figure}	

	Figures 1 (a) and (b)  show   the $T$-dependence of in-plane and out-of-plane surface impedance, $Z_s^{ab}=R_s^{ab}+iX_s^{ab}$ and $Z_s^c=R_s^c+iX_s^c$, respectively,  measured by Pb-resonator in the Meissner phase.  Both sets of data were obtained by  the measurements of $Z_s^{H_{\omega}\parallel c}$ and $Z_s^{H_{\omega}\parallel ab}$ (see the insets of Fig.~1).  In the configuration {\boldmath $H$}$_{\omega}\parallel c$, $Z_s^{ab}=Z_s^{H_{\omega}\parallel c}$ because the oscillatory  currents flow within the $ab$-plane. For {\boldmath $H$}$_{\omega}\parallel ab$, $Z_s^{H_{\omega}\parallel ab}$ is determined by the geometrical mean value of $Z_s^{ab}$ and $Z_s^c$ as  $Z_s^{H_{\omega}\parallel ab}=\frac{S^{ab}Z_s^{ab}+S^cZ_s^c}{S^{ab}+S^c}$.  Here $S^{ab}$ and $S^c$ are the areas of the faces where the screening currents flow in the $ab$-plane and along the $c$ direction, respectively \cite{kitano}.  Since $S^{ab}\sim 3S^{c}$,  and   $Z_s^c\sim 1.7 Z_s^{ab}$  from the anisotropy of  normal state resistivities ($\rho_n^c/\rho^{ab}_n\sim3$ \cite{elt}),   $Z_s^{c}$ and  $Z_s^{ab}$ make contribution in the same order to $Z_s^{H_{\omega}\parallel ab} $.   In both configurations,  both $R_s$ and $X_s$ decrease rapidly with decreasing $T$ below $T_c$.   In the Meissner phase,  microwave response is purely reactive and $R_s\simeq$ 0 and $X_s=\mu_0\omega \lambda$, where $\mu_0$ is the  permeability of  free space, $\omega/2\pi$ is the microwave frequency, and $\lambda$ is the London penetration length.  On the other hand,   response is dissipative in the normal state and $R_s=X_s=\mu_0\omega\delta_n/2$, where $\delta_n=\sqrt{2\rho_n/\mu_0\omega}$ is the skin depth and $\rho_n$ is the normal state resistivity.   The absolute value of $R_s^{i}$ and $X_s^{i}$ ($i=ab$ and $c$) were determined by comparison with dc resistivity $\rho_n^{i}$ assuming that $R_s^{i}=X_s^{i}=\sqrt{\mu_0\omega\rho_n^{i}/2}$ in the normal state.   Using $\rho_n^{ab}\sim 2 \mu \Omega$ cm and $\rho_n^c/\rho_n^{ab}\sim$3 at the onset,  the in-plane and out-of-plane penetration lengths, $\lambda_{ab}(T=0)\sim 500 \AA$ and $\lambda_{c}(T=0)\sim 750 \AA$, respectively,  were determined.  Our $\lambda_{ab}$ is smaller than the reported values \cite{hussey}, but the $H$-dependence of $Z_s$ is little affected by the absolute value of $\lambda_{ab}$.   The obtained value of $\lambda_{c}/\lambda_{ab}\sim1.5$ is much smaller than the anisotropy of upper critical fields, $H_{c2}^{ab}/ H_{c2}^{c}\simeq 4$ (see the inset of Fig.~2(b)), where $H_{c2}^{ab}$ and $H_{c2}^c$ are the upper critical field in {\boldmath $H$} parallel to the $ab$-plane and $c$-axis, respectively.   We note that the small anisotropy of the penetration lengths is consistent with the recent result  in Ref.\cite{cubit}.

\begin{figure}[t]
\begin{center}
\includegraphics[width=2.4in]{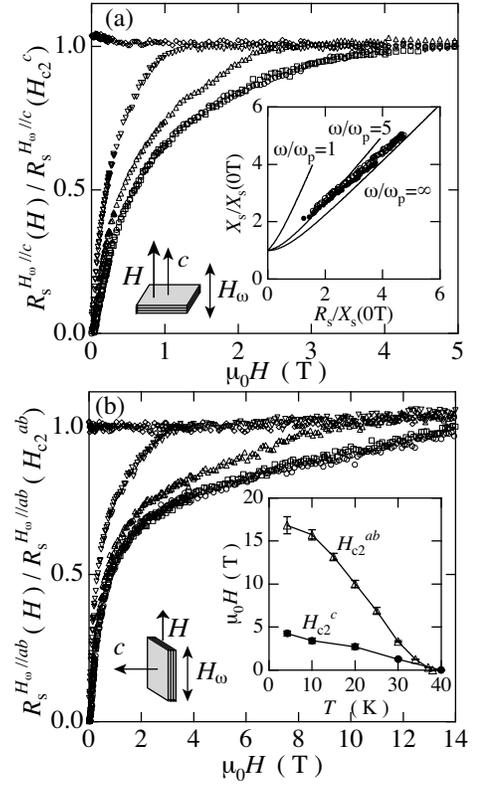}
\caption{(a) $H$-dependence of surface resistance $R_s^{H_{\omega} \parallel c}$ normalized by the normal state values measured in  {\boldmath $H$}$\parallel$ {\boldmath $H$}$_{\omega}\parallel c$ at 40~K (diamonds), 30~K (inverted triangles), 20~K (triangles), 10~K (squares), and 2.3~K (circles).  Inset:  A plot of  impedance  $R_s/X_s$(0T) vs $X_s/X_s$(0T) for   {\boldmath $H$}$\parallel$ {\boldmath $H$}$_{\omega}\parallel c$ (filled circles) and  for  {\boldmath $H$}$\parallel$ {\boldmath $H$}$_{\omega}\parallel ab$ (open circles).    The solid lines are the results calculated  from Eq.~(3) at different pinning frequencies.   (b) The same plot for  $R_s^{H_{\omega} \parallel ab}$ measured in  {\boldmath $H$} $\parallel$ {\boldmath $H$} $_{\omega} \parallel ab$.  Inset:  $T$-dependence of $H_{c2}^{ab}$ and $H_{c2}^c$, as determined by the field at which $R_s$ is the normal state value.  }
\end{center}
\end{figure}

	Figures 2 (a) and (b) show $H$-dependence of $R_s^{H_{\omega}\parallel c}$ ( {\boldmath $H$}$_{\omega}\parallel ${\boldmath $H$}$\parallel c$) and $R_s^{H_{\omega}\parallel ab}$ ( {\boldmath $H$}$_{\omega}\parallel ${\boldmath $H$}$\parallel ab$) measured by Cu-resonator.   In both configurations, $R_s$ increase rapidly with $H$.    In the vortex state, $Z_s$ is determined by  vortex dynamics.  According to Coffey and Clem, $Z_s$ can be  expressed as , 
\begin{equation}
Z_s=i\mu_0\omega\lambda\left[ \frac{1-(i/2)\delta_v^2/\lambda^2}{1+2i\lambda^2/\delta_{qp}^2}\right]^{1/2},
\label{eqn:CC}
\end{equation}	
where $\delta_v^2=\delta_f^2(1-i\omega_p/\omega)^{-1}$,  $\omega_p$ is the pinning frequency,  $\delta_f=\sqrt{2\rho_f/\mu_0\omega}$ is the FFF skin depth,  $\delta_{qp}=\sqrt{2\rho_{qp}/\mu_0\omega}$ with the quasi-particle resistivity $\rho_{qp}$ is the normal-fluid skin depth \cite{cc}.   To estimate the pinning frequency of MgB$_2$,  plots of $X_s$ vs $R_s$ in two configurations are shown in the inset of Fig.~2(a).  For comparison, the same data at several $\omega/\omega_p$ calculated from  Eq.(\ref{eqn:CC}) are also plotted.   The inset of Fig.~2(a) strongly suggests that  $\omega/\omega_p$ is much larger than unity in  both configurations, demonstrating that the  FFF state is nearly realized in the present frequency range.    In the FFF state, two characteristic length scales, namely $\lambda$ and  $\delta_f$, emerge in accordance with microwave penetration.   In  low fields, $\lambda$ greatly exceeds $\delta_f$($\lambda \gg \delta_f$).  In this region, $R_s$ and $X_s$ are given by $R_s\simeq\rho_f/\lambda$ and $X_s\simeq \mu_0\omega_\lambda$.  In high fields where $\delta_f$ exceeds $\lambda (\delta_f\gg\lambda)$, the response is similar to the normal state ($R_s\simeq X_s$), except that $\delta_n$ is replaced by $\delta_f$.  In MgB$_2$, this cross over occurs at a very low field ($\mu_0H<$~100~mT) in both configurations.  This indicates that the influence of penetration lengths on the vortex dynamics is negligibly small except at very low field regimes.

	Having established that the FFF state is realized in the present experiments,   FFF resistivities in  {\boldmath $H$} parallel to the $c$-axis and $ab$-plane, $\rho_f^{H\parallel c}$ and $\rho_f^{H\parallel ab}$, respectively,  are estimated next from Figs.2(a) and (b), assuming $\omega/\omega_p=\infty$.   Figure 3 (a)  shows $\rho_f^{H\parallel c}$  and $\rho_f^{H\parallel ab}$,  normalized by the normal state values as a function of $H/H_{c2}$ at $T$=2.3~K ($T/T_c$=0.06).    For  comparison,  the $H$-dependence of the FFF resistivity of single gap $s$-wave superconductor, as given by Eq.~(\ref{eqn:FFFS}),  and of $d$-wave Bi:2201, as reported in Ref.[\onlinecite{matsuda}]  are also shown.   The field dependence of both $\rho_f^{H\parallel c}$ and $\rho_f^{H\parallel ab}$ are spectacularly different from those of $s$-wave and $d$-wave superconductors.  Both $\rho_f^{H\parallel c}$ and $\rho_f^{H\parallel ab}$ increase steeply with $H$ at low fields, even steeper than $\rho_f$ in  $d$-wave superconductor.   In particular,  at $H/H_{c2}^{ab}\simeq0.1$, $\rho_f^{H\parallel ab}$ of MgB$_2$ is nearly half of $\rho_f^{H\parallel ab}$ at $H_{c2}$, indicating a striking enhancement of energy dissipation at very low fields.  This unusual FFF state of MgB$_2$ can be seen more clearly  in Fig.~3(b), in which the same data is plotted as a function of $\sqrt{H/H_{c2}}$.     A crossover field $H_{cr}$ which separates two distinct flux flow regime with different $\rho_f^{H\parallel ab}$ slope is distinguishable at $\sqrt{H_{cr}/H_{c2}^{ab}}\simeq 0.3$, as indicated by the arrow.  On the other hand,  such  a crossover cannot be easily identified in $\rho_f^{H\parallel c}$. 
	
\begin{figure}[t]
\begin{center}
\includegraphics[width=2.5in]{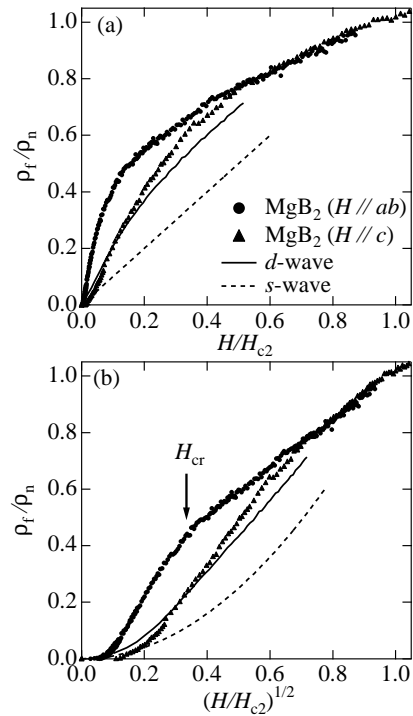}
\caption{(a) Flux flow resistivities $\rho_f^{H\parallel c}$ (solid triangles) and $\rho_f^{H\parallel ab}$ (solid circles) at 2.3~K normalized by the normal state values as a function of  $H$ normalized by the upper critical fields.   The solid line represents the FFF resistivity  of conventional $s$-wave superconductor expressed by Eq.~(1).  The dashed line represents  the FFF resistivity in $d$-wave superconductor reported in Ref.[\protect\onlinecite{matsuda}].  (b) The same data plotted as a function of $\sqrt{H/H_{c2}}$.  }
\end{center}
\end{figure}	
	
	 The appearance of the crossover field can be attributed to the two distinct superconducting gaps.   Unusual electronic structure in the vortex state of MgB$_2$ has been reported in several experiments.  According to STM measurements, the vortex radius is much larger than expected from $H_{c2}$ and superconductivity is strongly suppressed by a low field.  Sharp increases in the heat capacity $C_e$ \cite{hc} and  thermal conductivity $\kappa_e$ \cite{tc} at low fields with initial slopes much larger than in conventional $s$-wave superconductors also revealed the  unusual vortex state of MgB$_2$.    In particular, in {\boldmath $H$}$\parallel ab$, $C_e$ is nearly half of the normal state value $C_e^n$ at $H/H_{c2}^{ab}\sim0.1$, followed by a gradual increase up to $H_{c2}$.  Moreover, $\kappa_e$ in  {\boldmath $H$}$\parallel ab$ becomes nearly $H$-independent  after $\kappa_e$ reaches nearly half of the normal state value $\kappa_n$ at  $H/H_{c2}^{ab}\sim0.1$.   These strong enhancements  of $C_e$ and $\kappa_e$ at low fields has been attributed to the suppression of the small superconducting gap originating from the  three dimensional $\pi$-band.   Therefore,  it is natural to consider that {\it the steep increase and  crossover behavior observed in the FFF resistivities could  also be attributed to the two superconducting gaps.} 
	
	 In  conventional $s$-wave superconductors, the energy dissipation associated with  FFF is caused by the quasiparticles trapped within vortex cores \cite{kopnin,esc}.  Since in conventional $s$-wave superconductors the DOS within cores is approximately $H$-independent,   FFF resistivity is proportional to the number of  vortices as given by Eq.(\ref{eqn:FFFS}).   The situation in MgB$_2$ is very different.   Below $H_{cr}$,  quasiparticles are trapped by both small and large gaps.  Since the coherence lengths associated with the small gap are much larger than those associated with the large gap,  the number of quasiparticles trapped within cores well below $H_{cr}$ is much larger than the number trapped solely by  the large gap.   This gives rise to strong enhancement of  FFF resistivity at low fields.    As $H$ approaches $H_{cr}$, the small gap is strongly suppressed.  As a result,  the contribution of  bound quasiparticles of the $\pi$-band essentially becomes negligible above $H_{cr}$.  This  gives rise to a gentler  increase in FFF resistivity above $H_{cr}$ up to $H_{c2}$.   The kink structure observed at $H_{cr}$  in Fig.~3(b) implies that this crossover occurs in a narrow field range.  Similar to the $d$-wave superconductor described by Eq.(\ref{eqn:FFFD}), $\rho_f^{H\parallel ab}$  below and above $H_{cr}$ exhibit  nearly $\sqrt{H}$ dependence.   To clarify the origin of this $H$-dependence, more detailed information on quasiparticle structure and the dissipation mechanism is strongly required.

	Recently,   the carrier scattering rate in the $\pi$-band has been suggested \cite{mazin,raman} to be much higher than in the $\sigma$ band.    Detailed information about the quasiparticle structure and  scattering rate can be extracted when the heat capacity, thermal conductivity and  FFF resistivity data are combined.  In fact  heat capacity is essentially determined by the number of the quasiparticles.   On the other hand,  thermal conductivity is determined by  delocalized quasiparticles outside  vortex cores because  localized quasiparticles  cannot carry  heat.   In contrast,  energy dissipation in the FFF state is dominated by the quasiparticles  localized within cores.     We stress that all quantities,  heat capacity, thermal conductivity, and FFF resistivity in {\boldmath $H$}$\parallel ab$ follow  a striking similar trend;  a crossover from low field to high field region occurs at $H_{cr}/H_{c2}^{ab}\simeq 0.1$ and all quantities increase up to nearly half of the normal state values near $H_{cr}$ with  $C_e(H_{cr})/C_e(H_{c2}^{ab})\sim0.55$, $\kappa(H_{cr})/\kappa(H_{c2}^{ab})\sim0.57$, and $\rho_f^c(H_{cr})/\rho_f^c(H_{c2}^{ab})\sim0.45$.  This implies that {\it the carrier scattering rate  of the $\sigma$ band is close to that of $\pi$ the band}.  This appears to be  inconsistent with the suggestion by  Ref.[\onlinecite{mazin,raman}].    A more detailed microscopic calculation is needed to evaluate the carrier scattering rate.

	In summary,  the FFF resistivity in MgB$_2$ exhibits unusual field dependence, which is markedly different from conventional $s$-wave superconductors.   Two distinct  regimes in which the FFF resistivity exhibits  different  field dependence are  observed.   These unusual $H$-dependences indicate that  two very differently sized superconducting gaps manifest  in energy dissipation associated with the vortex dynamics.  
	
	We thank M.B.~Gaifullin, N.~Hayashi, Y.~Kato,  T.~Kita and E.B.~Sonin for valuable discussions. This work was supported by the New Energy and Industrial Technology Development Organization (NEDO) as collaborative research and development of fundamental technologies for superconductivity applications.

\end{document}